\documentstyle[11pt]{article}

\newfont{\msbm}{msbm10}
\def\gg{\hbox{\tt g}}
\def\CC{{\cal C}}

\def\P{{\cal P}}

\def\A{{\cal A}}

\def\R{\hbox{\msbm R}}

\newtheorem{prop}{Proposition}
\newtheorem{cor}{Corolary}
\newtheorem{theo}{Theorem}
\newtheorem{deff}{Definition}

\def\bc{\begin{cor}}
\def\ec{\end{cor}}

\def\bt{\begin{theo}}
\def\et{\end{theo}}

\def\bd{\begin{deff}}
\def\ed{\end{deff}}

\def\bp{\begin{prop}}
\def\ep{\end{prop}}

\def\ba{\begin{eqnarray}}
\def\ea{\end{eqnarray}}

\def\be{\begin{equation}}
\def\ee{\end{equation}}

\newfont{\msbms}{msbm6}  
\def\G{{\cal G}}

\def\Sig{\Sigma}

\def\Ab{\mbox{$\bar A$}}
\def\ab{\mbox{$\bar \A$}}
\def\Si{\mbox{$\Sigma$}}

\begin{document}

%%%%%%%%%%%%%%%%%%%%%%%%%%%%%%
%\baselineskip=18pt
%%%%%%%%%%%%%%%%%%%%%%%%%%%%%

%\markboth{J.M. Velhinho}
%{Functorial aspects of generalized connections}

%%%%%%%%%%%%%%%%%%%%%%%%%%%%%%%%%%%%%%%%%%%%%%%%%%%%%%%%%%%%%%%%%%%%%%
\title{Functorial Aspects	of the \\ Space of  Generalized Connections\footnote{Presented at the
{\it Workshop on Quantum Gravity
and Noncommutative Geometry}, Universidade Lus\'ofona, Lisbon, July 2004.}}

\author{J. M. Velhinho}

\date{{ Departamento de F\'\i sica, Universidade da Beira 
Interior}
\\R. Marqu\^es D'\'Avila e Bolama,
6201-001 Covilh\~a, Portugal\\jvelhi@dfisica.ubi.pt}

\maketitle

%%%%%%%%%%%%%%%%%%%%%%%%%%%%%%%%%%%%%%%%%%%%%
\begin{abstract}
\noindent 
We give a description of the category structure of the space of 
generalized connections, an extension of the space of connections that plays a central
role in loop quantum gravity.
\end{abstract}
%%%%%%%%%%%%%%%%%%%%%%%%%%%%%%%%%%%%%%%%%%%%%%
%%%%%%%%%%%%%%%%%%%%%%%%%%%%%%%%%%%%%%%%%%%%

%%%%%%%%%%%%%%%%%%%%% BEGIN PAPER %%%%%%%%%%%%%%%%%%%%%%%%%

%%%%%%%%%%%%%%%
\section{Introduction}
\label{int}
%%%%%%%%%%%%%%%%%%%%%%%%%%%%%%%%%%%%%%%%%
A central object in loop quantum gravity~\cite{ALR,R2,T1} is the so-called kinematical
Hilbert space. This is the space where basic kinematical operators, as well as quantum versions
of the constraints of canonical general relativity, are defined. The kinematical Hilbert space  is an $L^2$
space of functions on a compact space \ab~\cite{AI,AL1,B1}, square integrable with respect
to the Ashtekar-Lewandowski measure~\cite{AL1}.
Properties of the space \ab, known as the space of generalized connections, therefore play an
important role in loop quantum gravity.

In this contribution we focus on one particular aspect of the space \ab.
As first pointed out by Baez~\cite{B3},
the proper framework to express the algebraic 
properties of \ab\ implicit in earlier formulations~\cite{B1,AL3,ALMMT} uses the language of category
theory. Following previous works~\cite{Ve,VR}, we present here a concise description of the category
structure of \ab. In section \ref{sec2} we present the space \ab\ as a space of functors equipped 
with a natural topology. In section \ref{gauge} 
we show how extensions of the group of gauge transformations and of the group of diffeomorphisms 
fit within the category formulation of \ab.

%%%%%%%%%%%%%%%%%%%%%%%%%
%%%%%%%%%%%%%%%%%%%%%%%%%%%%%  Section  %%%%%%%%%%%%%%%%%%%%%%%%%%%%%%%%%
\section{Generalized connections}
\label{sec2}
%%%%%%%%%%%%%%%%%
%%%%%%%%%%%%%%%
\subsection{Groupoid of paths $\P$}
\label{seg}
%%%%%%%%%%%%%%%%%%%%%%%%%%%%%%%%%%%%%%%%%%%%%%%%%%%%%%%%%%%%%%%%%%%%%%%%%
%%%%%%%%%%%%%%%%%%%%%%%%%%%%%%%%%%%%%%%%%
Let \Si\ be an analytic, connected and orientable $d$-dimensional manifold. 
Let us consider the set $\CC$ of all continuous, oriented and piecewise analytic
parametrized curves in \Si , i.e.~maps
%%%%%%%%%%%%%
$$
c:[0,t_1]\cup\ldots\cup [t_{n-1},1]\to\Sig
$$
%%%%%%%%%%%%
that are continuous in all the domain $[0,1]$, analytic in the closed
intervals $[t_k,t_{k+1}]$ and such that the images $c\bigl(]t_k,t_{k+1}[
\bigr)$ of the open intervals $]t_k,t_{k+1}[$ are submanifolds
embedded in \Si . Maps $s$  and $r$  are
defined, respectively, by $s(c)=c(0)$, $r(c)=c(1)$. 

Given two curves
$c_1,c_2\in\CC$ such that $s(c_2)=r(c_1)$, let $c_2c_1\in\CC$
denote the natural composition given by
%%%%%%%%%%%
$$
(c_2c_1)(t)=\left\{\begin{array}{lll} c_1(2t), & {\rm for} & t\in[0,1/2] \\
c_2(2t-1), & {\rm for} & t\in[1/2,1]\,. \end{array} \right.
$$
%%%%%%%%%%%
Consider also the operation $c\mapsto c^{-1}$ 
given by $c^{-1}(t)=c(1-t)$. The above composition
of parametrized curves is not truly associative, since the curves $(c_3c_2)c_1$ 
and $c_3(c_2c_1)$ are related by a reparametrization, i.e.~by an
orientation preserving piecewise analytic diffeomorphism $[0,1]\to [0,1]$.
Similarly, the curve $c^{-1}$ is not the inverse of the curve $c$.
(We refer to compositions of the form $c^{-1}c$ as retracings.)

To achieve a well defined associative composition, with inverse, one considers
the following equivalence relation:
two curves $c,c'\in\CC$ are said to be equivalent if
$s(c)=s(c')$, $r(c)=r(c')$ and
$c$ and $c'$ coincide up to a finite number of retracings
and a reparametrization.

We   denote the set of
the above defined equivalence classes by $\P$, and refer to generic elements
of $\P$ as paths $p$.

The set of paths $\P$ is then naturally equipped with a groupoid structure, as follows.
The composition of paths
is defined by the composition of elements of $\CC$:  if $p,p'\in\P$ 
are such that $r(p)=s(p')$, one defines $p'p$ as the 
equivalence class of $c'c$, where $c$ (resp.~$c'$) belongs to the class
$p$ ($p'$). The composition 
in $\P$ is now associative,
since $(c_3c_2)c_1$ and $c_3(c_2c_1)$ belong to the same equivalence class.
The points of $\Si$ play the role of objects in this context.
Points are in one to one correspondence with identity
paths:  given $x\in\Sig$, the corresponding identity ${\bf 1}_x\in\P$ is the
equivalence class of $c^{-1}c$, with $c\in\CC$ such that $s(c)=x$. 
If $p$ is 
the class of $c$ then $p^{-1}$ is the class of $c^{-1}$. It is clear that
$p^{-1}p={\bf 1}_{s(p)}$ and 
$p p^{-1}={\bf 1}_{r(p)}$. 

One, therefore, has a well defined 
groupoid, whose set of objects is $\Si$ and whose set of arrows is $\P$.
As usual, we will use the same notation, $\P$, both for
the set of arrows and for the groupoid.
%%%%%%%%%%%%%%%%%%%%%%%%%%
%%%%%%%%%%%%%%%%%%%%%%%%%%% Section %%%%%%%%%%%%%%%%%%%%%%%%%%%%%%%%%%%%%%%%
\subsection{Set of functors ${\rm Hom}\,[\P,G]$}
\label{ssi}
%%%%%%%%%%%%%%%%%%%%%%%%%%%%%%%%%%%%%%%%%%%%%%%%%%%%%%%%%%%%%%%%%%%%%%%%%%%%
%%%%%%%%%%%%%%%%%%%%%%%%%%%%%%%%
Let $G$ be a (finite dimensional) connected and compact
Lie group. Let
${\rm Hom}\,[\P,G]$ denote the set of 
functors from the groupoid $\P$ to the group $G$, i.e.~the set of 
maps $\bar A:\P\to G$ such that $\bar A(p'p)=\bar A(p')\bar A(p)$ and
$\bar A(p^{-1})=\bar A(p)^{-1}$.

Let us 
show that the space $\A$ of smooth $G$-connections on any given principal 
$G$-bundle over $\Si$ is a subset of ${\rm Hom}\,[\P,G]$.
We  assume that
a fixed trivialization of the bundle  has been chosen. Connections
can then be identified with local connection potentials.

The space of connections $\A$ is injectively mapped into ${\rm Hom}\,[\P,G]$ through the
use of the parallel transport, or holonomy, functions. The holonomy
defined by a  connection $A\in\A$ and a  curve $c\in\CC$
is denoted by $h_c(A)$. Using the fixed trivialization, one can assume that
holonomies $h_c(A)$ take values on the group $G$. The following properties
of holonomies  hold: 
{\it i}\/) $h_c(A)$ is invariant under 
reparametrizations of $c$; {\it ii}\/) $h_{c_1 c_2}(A)=h_{c_1}(A)h_{c_2}(A)$ 
and 
{\it iii}\/) $h_{c^{-1}}(A)={h_c(A)}^{-1}$. One thus conclude that 
$h_c(A)$ depends only
on the equivalence class of $c$, i.e., for fixed $A\in\A$, $h_c(A)$ defines a function
on $\P$, with values on $G$. This function is, moreover, a functor, by {\it ii}\/).
Summarizing, we have a map $\A\to {\rm Hom}\,[\P,G]$:
%%%%%%%%%%%%%%%%%%%%
\be
\label{image}
A\mapsto \bar A_A,\ \bar A_A(p):=h_p(A),\ \forall p\in\P.
\ee
%%%%%%%%%%%%%%%    
That this map is injective is guaranteed by the  fact that the set
of holonomy functions $\{h_c,\ c\in\CC\}$ separates points in 
$\A$~\cite{G},
i.e.~given $A\not =A'$ one can find $c\in\CC$ such that $h_c(A)\not = h_c(A')$.

The set ${\rm Hom}\,[\P,G]$ is, however, much larger than  $\A$.
To begin with, depending on $\Si$ and $G$, different bundles may exist, and
${\rm Hom}\,[\P,G]$ contains the space of connections of all these bundles.
Moreover, elements of ${\rm Hom}\,[\P,G]$ that do not correspond 
to any smooth connection do exist~\cite{AI,AL1}.

%%%%%%%%%%%%%%%%%%%%%
%%%%%%%%%%%%%%%%%%%%%%%%%%%%%%%%%%
\subsection{Topology}
\label{pro}
%%%%%%%%%%%%%%%%%%%%%%%%%%%%
%%%%%%%%%%%%%%%%%%%%%%%%%%

A natural topology in ${\rm Hom}\,[\P,G]$ can be introduced in several equivalent  ways. 
The one more closely related to well known methods in quantum field theory
uses the projective structure of ${\rm Hom}\,[\P,G]$~\cite{AL1,MM,AL2}. We follow a
different, more direct, approach.

Let us consider the set of {\it all} maps from $\P$ to $G$, identified with the product space 
$\times_{p\in\P}G$. The product space is compact Hausdorff when equipped with the Tychonov topology. 
We thus let the topology 
on ${\rm Hom}\,[\P,G]\subset\times_{p\in\P}G$ be the subspace topology. It is clear that ${\rm Hom}\,[\P,G]$
then becomes a compact Hausdorff space:
the Hausdorff property  is inherited from $\times_{p\in\P}G$, and from the
fact that  ${\rm Hom}\,[\P,G]$ contains only  functors follows easily that it is a closed set,
therefore compact.

A more explicit characterization of the topology on ${\rm Hom}\,[\P,G]$ is the following:
it is the 
weakest topology such that all maps
%%%%%%%%%%%%%%
\be
\label{6}
\pi_p: {\rm Hom}\,[\P,G] \to G,\ \ \bar A\mapsto \bar A(p),
\ee
%%%%%%%%%%%%%%%%
are continuous, $\forall   p\in\P$.  
The thus obtained compact Hausdorff space is the so-called space of generalized connections,
usually denoted by $\bar\A$.

%%%%%%%%%%%%%%%%
%%%%%%%%%%%%%%%%%%%%%%%%
\section{Generalized gauge transformations and diffeomorphisms}
\label{gauge}
%%%%%%%%%%%%%%%%%%%%%%%%%%%%%%
%%%%%%%%%%%%%%%%
Two  groups act naturally and continuously on $\bar\A$. 
One  is the group 
of natural transformations  of the set of functors ${\rm Hom}[\P,G]\equiv\ab$. 
This group is well understood,
and is widely accepted as the  generalization of the group
$\G=C^{\infty}(\Sigma,G)$ of smooth local gauge transformations.
The second group of interest is the group ${\rm Aut}(\P)$ 
of automorphisms
of the groupoid $\P$. It contains the group 
${\rm Diff}^{\omega}(\Sigma)$ of analytic diffeomorphisms as a subgroup.
Although extensions of ${\rm Diff}^{\omega}(\Sigma)$ are welcome in applications 
of the current formalism to quantum gravity, the potencial role of the group 
${\rm Aut}(\P)$, or subgroups thereof, is still not clear.

We begin by discussing the group of natural transformations of \ab.
In this case, natural transformations  
form the group,  denoted by $\bar\G$, of all maps $\gg:\Sigma\to G$, under 
pointwise multiplication.  The action of $\bar \G$ on $\bar \A$ can be written as
$\bar \A\times\bar \G\ni(\Ab,\gg)\mapsto {\Ab}_{\gg}$ such that
%%%%%%%%%%%%%%%%
\be
\label{3}
{\Ab}_{\gg}(p)=\gg(r(p))\Ab(p)\gg(s(p))^{-1}.
\ee
%%%%%%%%%%%%%%%%%%%%%
This action is readily seen to be continuous. In fact, since the topology on 
$\bar \A$ is the 
weakest  such that all maps $\pi_p$ (\ref{6}) are continuous, one can conclude
that a map
$\varphi:\ab\to\ab$ is continuous if and only if the maps $\pi_p\circ\varphi$
are continuous $\forall p$, which is  the case for elements of $\bar \G$.

Expression (\ref{3}) is a generalization of the action of smooth gauge 
transformations on the set of parallel transport functions $h_p(A)$ for smooth 
connections. It is therefore natural to accept $\bar \G$ as the generalized group
of gauge transformations on $\bar \A$. This extension of the gauge group is in fact 
required: since $\bar \A$ now contains arbitrary functors
from $\P$ to $G$, to mod out only by smooth gauge transformations would 
leave spurious degrees of freedom untouched~\cite{B1,AL3,Ve}.

Let us now turn to  the group ${\rm Aut}(\P)$ of automorphisms
of the groupoid $\P$. By definition, an automorphism of $\P$ is an invertible functor
from $\P$ to itself. An element $F\in {\rm Aut}(\P)$ is therefore characterized
by a bijection of $\Sigma$ and a composition preserving bijection on the set of paths,
such that $F({\bf 1}_x)={\bf 1}_{F(x)}$, $\forall x\in\Sigma$. The action of 
${\rm Aut}(\P)$ on $\bar \A$ is given by
%%%%%%%%%%%%%%%%
\be
\label{4}
\Ab\mapsto F\Ab\, :\  F\Ab(p)=\Ab(F^{-1}p),\ \forall p\in\P, \ F\in {\rm Aut}(\P).
\ee
%%%%%%%%%%%%%%%
The continuity of this action is clear, since $\pi_p\circ F=\pi_{F^{-1}p}$.

The group ${\rm Aut}(\P)$ contains as a subgroup the natural representation of
the group ${\rm Diff}^{\omega}(\Sigma)$ of analytic diffeomorphisms of $\Sigma$,
whose action on curves factors through the equivalence relation
that defines $\P$.
The group ${\rm Aut}(\P)$ therefore emerges, in the current  context, as the 
largest possible
extension of ${\rm Diff}^{\omega}(\Sigma)$. Extensions of
${\rm Diff}^{\omega}(\Sigma)$ are welcome in applications to quantum gravity. 
In fact, the very formalism suggests some sort of extension of 
${\rm Diff}^{\omega}(\Sigma)$. Like in the above mentioned replacement of the classical
gauge group $\G$ by $\bar \G$, such extension is not fully motivated from the classical
perspective, but it is, nevertheless, likely to be required for an appropriate 
implementation of diffeomorphism invariance, once $\A$ is replaced by $\bar \A$.

Several subgroups of ${\rm Aut}(\P)$ have been proposed so far as
candidates to an appropriate extension of ${\rm Diff}^{\omega}(\Sigma)$.
One such extension, including homeomorphisms of $\Sigma$ of class $C^n$, appears
in a recent report by Ashtekar and Lewandowski~\cite{ALR}. 
Another recent proposal is due to Fleischhack~\cite{Fl1}, who considers
the so-called stratified analytic diffeomorphisms.
Previous approaches allowed for stronger deviation from classical smoothness
(on lower dimensional subsets of $\Sigma$ only), like e.g.~in Zapata's~\cite{Z}
proposal of replacing ${\rm Diff}^{\omega}(\Sigma)$ by the  group
of piecewise analytic diffeomorphisms. (See also~\cite{R2,FR} for related work,
in the framework of piecewise smooth curves.)

%%%%%%%%%%%%%%%  Acknowledgments %%%%%%%%%%%%%%%%%%%%%%

\section*{Acknowledgements}
\noindent 
I thank the organizers of the Lus\'ofona Workshop on Quantum Gravity and Noncommutative
Geometry,
Aleksandar Mikovic, Nuno Costa Dias and Jo\~ao Nuno Prata. 
I am most
grateful to Jos\'e Mour\~ao  for  discussions.
This work was supported in part by 
%projects 
POCTI/33943/MAT/2000,
%, CERN/P/FIS/43171/2001, 
POCTI/FNU/\-49529/2002 
and POCTI/FP/FNU/50226/2003.
%%%%%%%%%%%%%%%%%%%

%%%%%%%%%%%%%% Bibliography %%%%%%%%%%%%%%%%%%%%%%%

%%%%%%%%%%%%%%%%%%%%%%%%%%%%%%%%%

\end{document}